\documentclass{elsart}

\usepackage{epsfig,amssymb}

\begin{document}

\begin{frontmatter}

\title{Search for Point-Like Sources of Cosmic Rays with Energies above
       $10^{18.5}$ eV in the HiRes--I Monocular Data-Set}
\vspace{-1.0cm}
\author[utah]{R.U.~Abbasi},
\author[utah]{T.~Abu-Zayyad},
\author[lanl]{J.F.~Amann},
\author[utah]{G.~Archbold},
\author[utah]{R.~Atkins},
\author[adel]{J.A.~Bellido},
\author[utah]{K.~Belov},
\author[mont]{J.W.~Belz},
\author[colu]{S.~BenZvi},
\author[rutg]{D.R.~Bergman},
\author[utah]{G.W.~Burt},
\author[utah]{Z.~Cao},
\author[adel]{R.W.~Clay},
\author[colu]{B.M.~Connolly},
\author[adel]{B.R.~Dawson},
\author[utah]{W.~Deng},
\author[utah]{Y.~Fedorova},
\author[utah]{J.~Findlay},
\author[colu]{C.B.~Finley},
\author[utah]{W.F.~Hanlon},
\author[lanl]{C.M.~Hoffman},
\author[lanl]{M.H.~Holzscheiter},
\author[rutg]{G.A.~Hughes},
\author[utah]{P.~H\"{u}ntemeyer},
\author[utah]{C.C.H.~Jui},
\author[utah]{K.~Kim},
\author[mont]{M.A.~Kirn\corauthref{cor1}},
\author[utah]{E.C.~Loh},
\author[utah]{M.M.~Maestas},
\author[toky]{N.~Manago},
\author[lanl]{L.J.~Marek},
\author[utah]{K.~Martens},
\author[nmex]{J.A.J.~Matthews},
\author[utah]{J.N.~Matthews},
\author[colu]{A.~O'Neill},
\author[lanl]{C.A.~Painter},
\author[rutg]{L.~Perera},
\author[utah]{K.~Reil},
\author[utah]{R.~Riehle},
\author[nmex]{M.D.~Roberts},
\author[toky]{M.~Sasaki},
\author[rutg]{S.R.~Schnetzer},
\author[adel]{K.M.~Simpson},
\author[lanl]{G.~Sinnis},
\author[utah]{J.D.~Smith},
\author[utah]{R.~Snow},
\author[utah]{P.~Sokolsky},
\author[colu]{C.~Song},
\author[utah]{R.W.~Springer},
\author[utah]{B.T.~Stokes},
\author[utah]{J.R.~Thomas},
\author[utah]{S.B.~Thomas},
\author[rutg]{G.B.~Thomson},
\author[lanl]{D.~Tupa},
\author[colu]{S.~Westerhoff},
\author[utah]{L.R.~Wiencke},
\author[rutg]{A.~Zech} \\

\vspace{0.5 cm}
\hspace{-0.3cm}
\author{\bf The High--Resolution Fly's Eye Collaboration}

\corauth[cor1]{Corresponding author. {\em E-mail Address}:
               mkirn@physics.montana.edu}

\address[utah]{University of Utah,
Department of Physics and High Energy Astrophysics Institute,
Salt Lake City, Utah, USA}

\address[lanl]{Los Alamos National Laboratory,
Los Alamos, NM, USA}

\address[adel]{University of Adelaide, Department of Physics,
Adelaide, South Australia}

\address[mont]{University of Montana, Department of Physics and Astronomy,
Missoula, Montana, USA.}

\address[rutg]{Rutgers --- The State University of New Jersey,
Department of Physics and Astronomy,
Piscataway, New Jersey, USA}

\address[toky]{University of Tokyo,
Institute for Cosmic Ray Research,
Kashiwa, Japan}

\address[nmex]{University of New Mexico,
Department of Physics and Astronomy,
Albuquerque, New Mexico, USA  }

\address[colu]{Columbia University, Department of Physics and
Nevis Laboratory, New York, New York, USA}

\title{}

\begin{abstract}
  We report the results of a search for point-like deviations from
  isotropy in the arrival directions of ultra--high energy cosmic
  rays in the northern hemisphere. In the monocular data set
  collected by the High--Resolution
  Fly's Eye, consisting of 1,525 events with energy exceeding
  $10^{18.5}$ eV, we find no evidence for point-like excesses.  We
  place 90\% c.l. upper limits less than or equal to 0.8 cosmic
  rays/km$^2$yr on the flux from such sources as a function of position
  in the sky.
\end{abstract}


\end{frontmatter}

\section{Introduction}
\label{sec-intro}

In the search for the origins of ultra--high energy cosmic rays
(UHECR), a variety of strategies have been employed to detect
deviations from isotropy in event arrival directions. The results
of these searches have been ambiguous. The most recent
results~\cite{dipolepaper} have supported the null hypothesis for
large--scale dipole behavior in arrival directions for particles
above $10^{18.5}$ eV.  However, previous
experiments~\cite{hayashida99} have found evidence for a dipole
moment in Right Ascension (RA) at energies above $10^{18}$ eV. On
smaller angular scales, excesses have been alternately claimed and
refuted in the vicinity of
Cygnus~X-3~\cite{cyg-flyseye,cyg-akeno,cyg-haverah,cyg-casa}, an
x-ray binary within our galaxy, including the report of a possible
excess in a point-source search~\cite{doi95}.

Point-like excesses at these energies can arise from only a
limited number of source scenarios. Galactic and extragalactic
magnetic fields are expected to produce large perturbations in the
arrival directions of charged particles: A proton of energy
$10^{18.5}$~eV may be deflected by several tens of degrees as it
traverses the disk of the Milky Way galaxy, with a typical
magnetic field of order 1~microgauss~\cite{galaxy}. A compact
arrival direction excess at these energies would therefore suggest
neutral primaries.  Neutrons however possess a lifetime of $3
\times 10^{12}$ seconds at $10^{18.5}$~eV, and therefore cannot
have originated more than 30 kpc from Earth. Thus any viable
source of standard model neutral hadronic matter would have to be
located within the Milky Way Galaxy.

In this paper, we conduct a search for point-like behavior in
arrival direction of cosmic ray events above $10^{18.5}$~eV in the
northern hemisphere, using a skymap technique in which we evaluate
our sensitivity using Monte Carlo simulated sources. In addition,
we consider the historically significant source candidate
Cygnus~X-3 as the focus of an {\em a priori} search.

\section{The HiRes Monocular Data Set}
\label{sec-dataset}

The High--Resolution Fly's Eye (HiRes) consists of two nitrogen
fluorescence observatories --- HiRes--I and HiRes--II ---
separated by 12.6 km and located at Dugway, Utah. HiRes was
conceived as a stereo detector, however due to the larger
available statistics it is desirable to reconstruct extensive
airshowers in monocular mode as well. This HiRes--I monocular data
set consists of 2,820 good--weather detector hours of data,
collected between May 1997 and February 2003. A total of 1,525
events with energies exceeding $10^{18.5}$ eV were collected
during this time and are included in the present analysis.

\begin{figure}[h]
   \epsfxsize=12.0cm
   \centerline{\epsffile{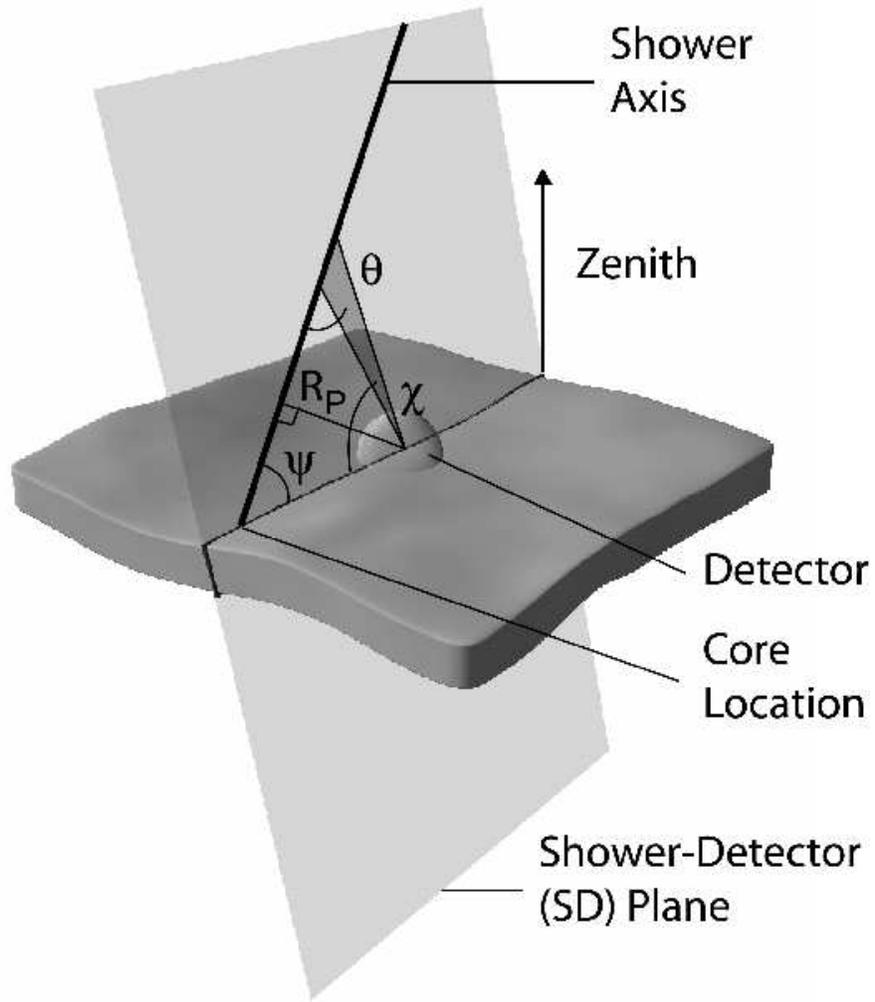}}
   \caption{\label{monogeom} The geometry of reconstruction for a
            monocular air fluorescence detector.}
\end{figure}

The HiRes-I monocular data set and airshower reconstruction by
profile constrained fitting has recently been described in the
literature~\cite{monopaper}.  A residual effect of the
profile--constrainted fit technique is orientation--dependent
(elliptical) uncertainties in the airshower arrival directions. In
Figure~\ref{monogeom}, the airshower reconstruction geometry is
illustrated for a monocular air fluorescence detector. In this
view, the shower--detector plane (SDP) for HiRes--I events is
well--reconstructed, with uncertainty parameterized as
\begin{equation}
\label{eq-sdp} \sigma_{SDP} =
88.151^{\circ}e^{-.51\Delta\chi}+0.374^{\circ}
\end{equation}
where $\Delta\chi$ is the angular tracklength of the shower in
degrees. Typical values of $\sigma_{SDP}$ for this analysis range
from $0.4^{\circ} \rightarrow 1.7^{\circ}$.  The angle of the
track within the SDP, $\Psi$, is less well reconstructed and is
parameterized by
\begin{equation}
\label{eq-psi} \sigma_{\Psi} =
18.381^{\circ}e^{-\log_{10}(1.45E)}+4.073^{\circ}
\end{equation}
where the energy $E$ is expressed in EeV ($10^{18}$~eV). Typical
values of $\sigma_{\Psi}$ in this analysis range from
$5.4^{\circ} \rightarrow 15^{\circ}$.

\begin{figure}[h]
   \epsfxsize=18.0cm
   \centerline{\epsffile{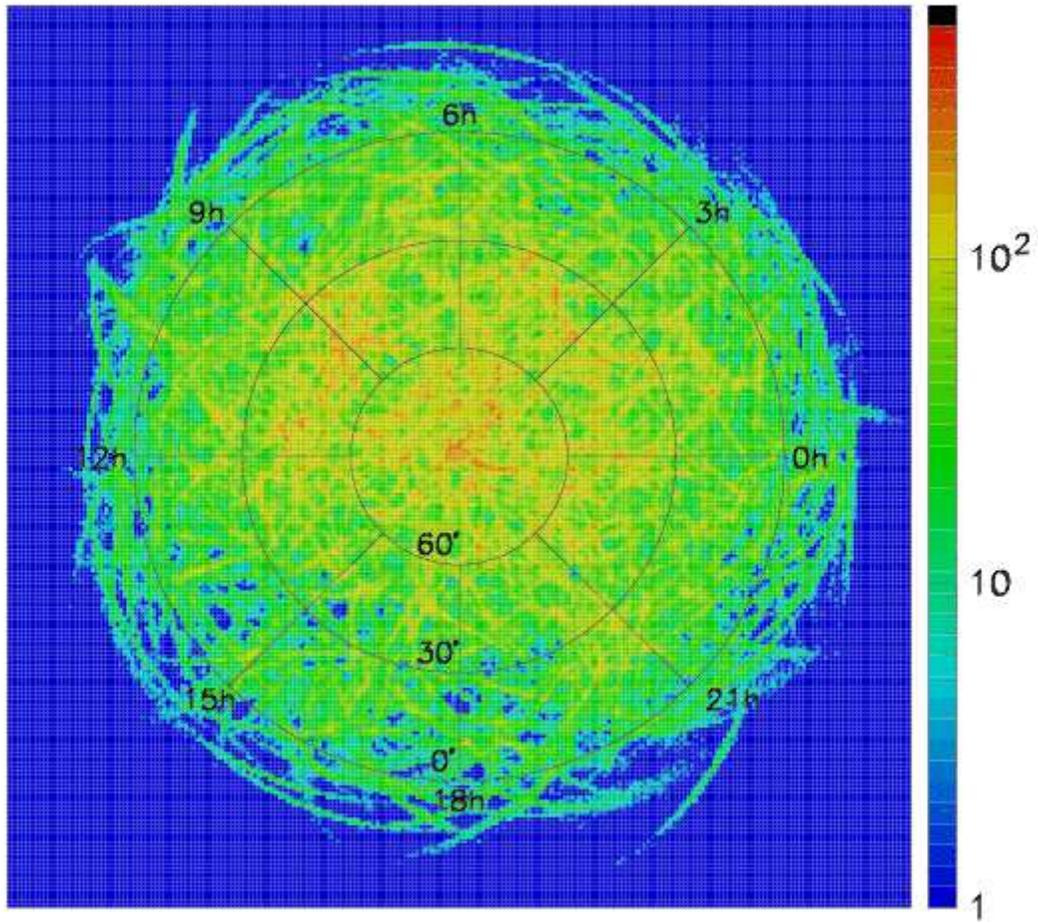}}
   \caption{\label{skymap} Skymap of arrival directions of events in the
    the HiRes--I monocular data set, plotted in polar projection, equatorial
    coordinates. Each HiRes event is represented by 1,000 points randomly
    thrown according to the elliptical Gaussian error model of
    Equations~\ref{eq-sdp} and \ref{eq-psi}.  The bin size in this
    plot (and all similar plots) is approximately $1^{\circ}\times1^{\circ}$.}
\end{figure}

In Figure~\ref{skymap}, we plot the skymap formed from the arrival
directions of events in the present data set. Each event's ``error
ellipse'' is represented by generating 1,000 points per event,
distributed according to the Gaussian error model of
Equations~\ref{eq-sdp} and \ref{eq-psi}.  Figure~\ref{skymap} is
plotted in equatorial coordinates as a polar plot.
Note that bins are assigned using a cartesian projection of the polar
plot shown in Figure~\ref{skymap} and all similar figures. As such,
angular bin size varies across the map, but averages
approximately $1^{\circ}\times1^{\circ}$.

We next discuss the Monte Carlo technique by which we evaluate the
significance of fluctuations in the skymap as well as our
sensitivity to point-like behavior in arrival direction.

\clearpage

\section{The Monte Carlo; Comparison of Data to Expectation from an
         Isotropic Background}
\label{sec-monte}

We use a library of simulated events, generated by the Monte Carlo
technique and reconstructed using the profile--constrained
reconstruction program to determine the background expectation for
isotropically distributed sources as well as to evaluate our
sensitivity to point-like behavior in arrival direction. For this
library, an isotropic distribution is assumed for events
possessing the spectrum and composition reported by the stereo
Fly's Eye experiment~\cite{bird93,bird94}.

\begin{figure}[h]
   \epsfxsize=15.0cm
   \centerline{\epsffile{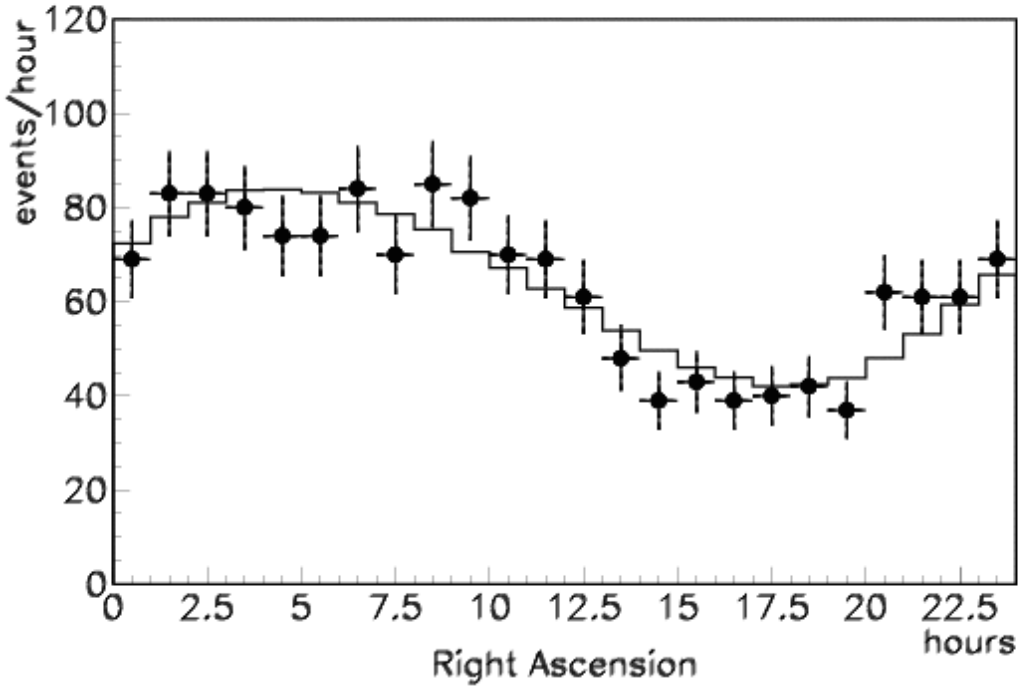}}
   \caption{\label{datamc_ra} Comparison of HiRes--I data (points)
   and Monte Carlo (solid histogram) distributions in right ascension
   (RA).}
\end{figure}

\begin{figure}[h]
   \epsfxsize=15.0cm
   \centerline{\epsffile{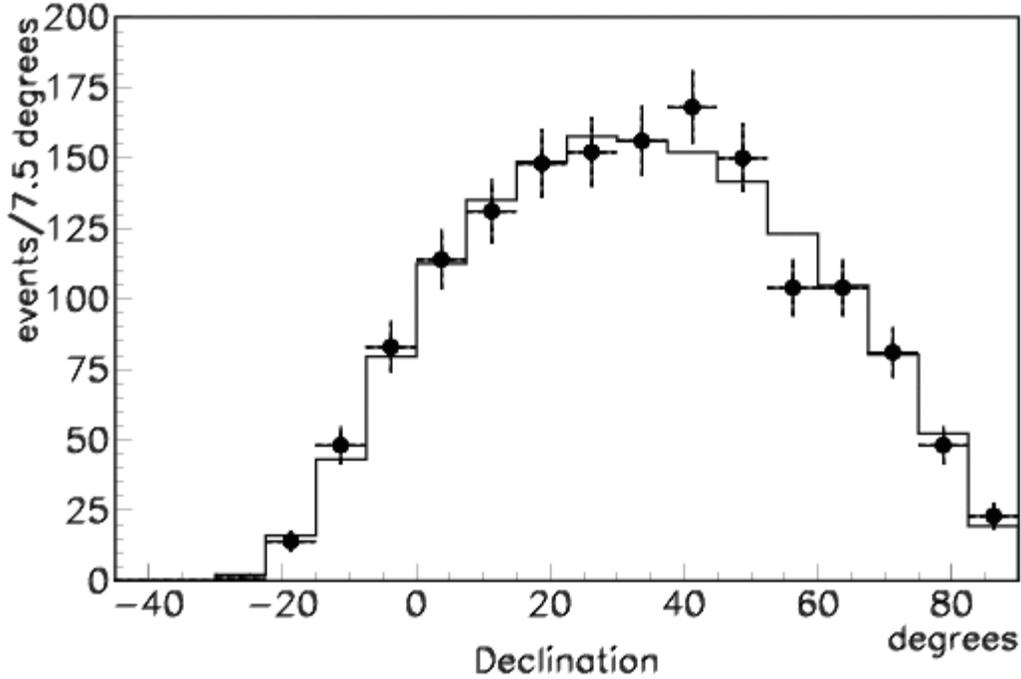}}
   \caption{\label{datamc_dec} Comparison of HiRes--I data (points)
   and Monte Carlo (solid histogram) distributions in declination (DEC).}
\end{figure}

A detector runtime database is used to randomly assign a time from
detector ``on'' periods to each event in the isotropic background
data set. A total of 1,000 isotropic data sets with the same sky
exposure as the HiRes--I monocular data set were generated for
comparison studies. Further discussion of this Monte Carlo can be
found in Reference~\cite{dipolepaper}. In Figures~\ref{datamc_ra}
and \ref{datamc_dec} we compare the data and Monte Carlo
distributions of events in the variables RA and DEC, respectively.

\begin{figure}[h]
   \epsfxsize=18.0cm
   \centerline{\epsffile{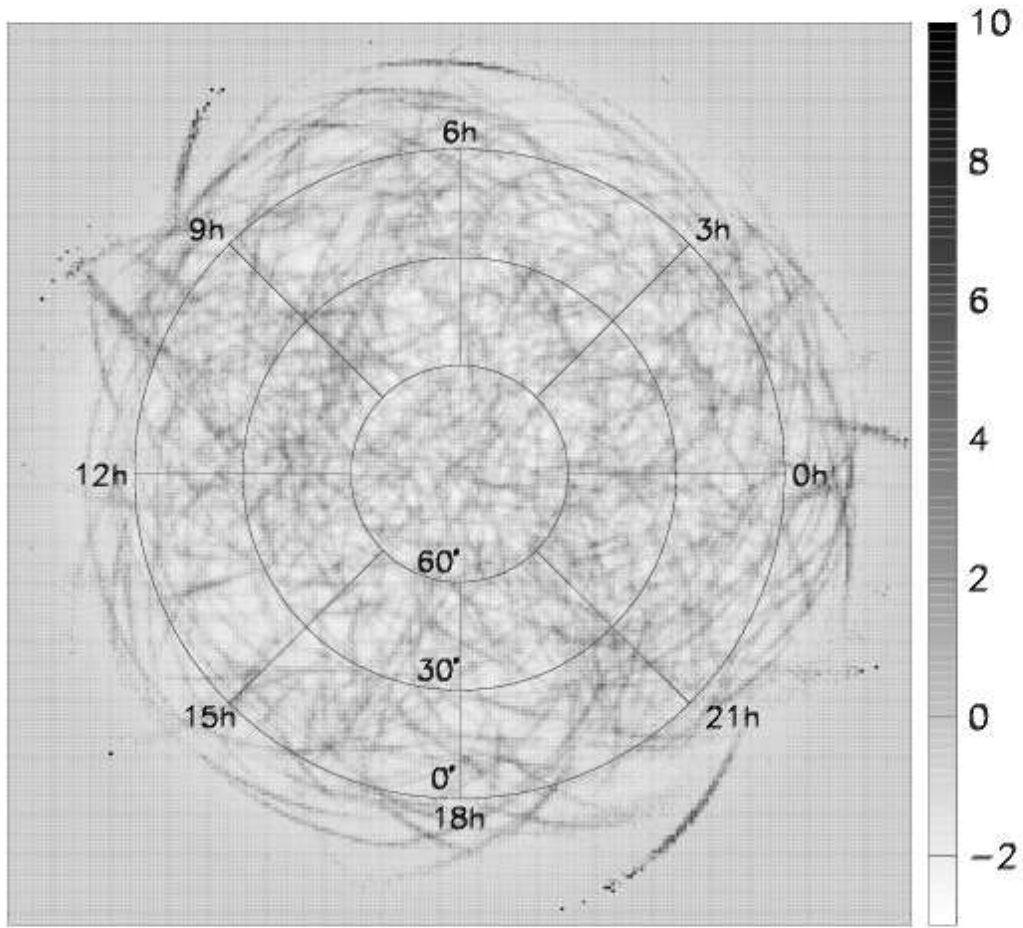}}
   \caption{\label{realchi} $\chi_1$ (Equation~\ref{eq-chione})
   distribution for the HiRes-I monocular data set.}
\end{figure}

In order to understand the significance of the fluctuations in
Figure~\ref{skymap}, we compare the data on a bin-by-bin basis to
the 1,000 simulated data sets. Defining $N_{DATA}$ as the bin
density of the data, $N_{MC}$ as the bin density of the simulated
isotropic data sets, and $\sigma_{MC}$ as the standard deviation
of the Monte Carlo bin density, the variable
\begin{equation}
\label{eq-chione}
\chi_1 = \frac{(N_{DATA} - \left< N_{MC} \right> )}{\sigma_{MC}}
\end{equation}
provides a measure of the fluctuation per bin.
Figure~\ref{realchi} shows the distribution of $\chi_1$ as a
function of position in the sky for the HiRes--I monocular data
set as extracted from this technique.

\begin{figure}[h]
   \epsfxsize=15.0cm
   \centerline{\epsffile{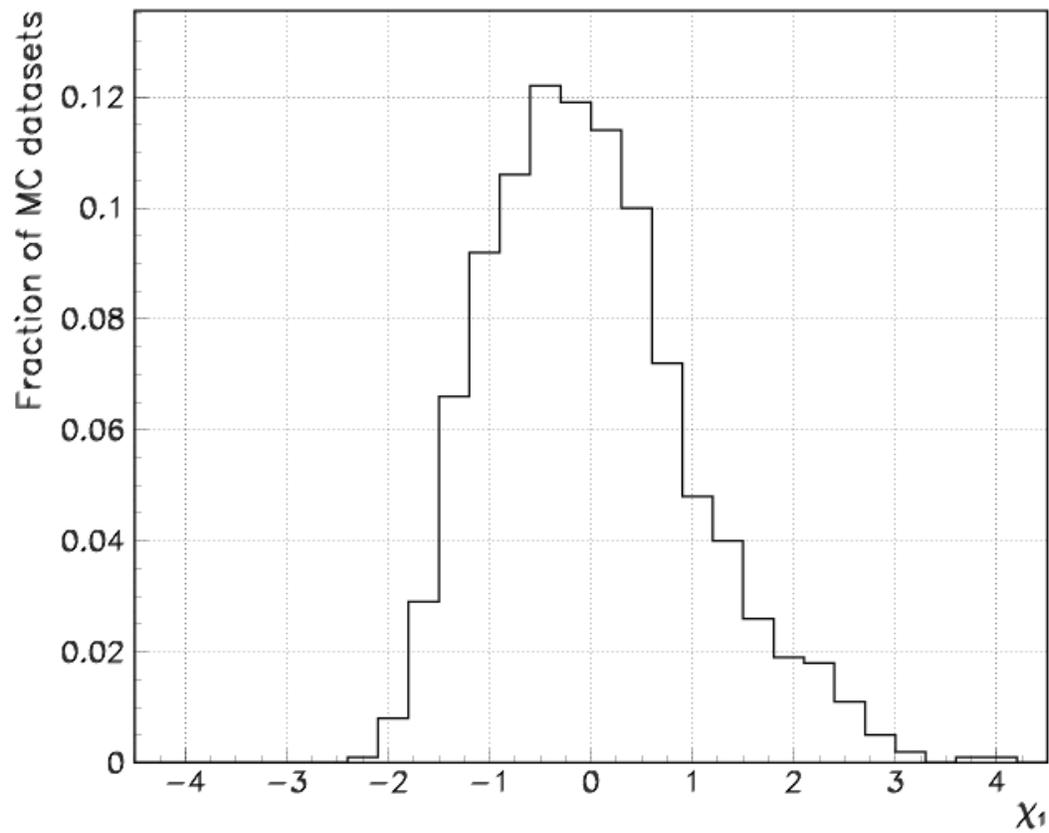}}
   \caption{\label{distrib} Example distribution of $\chi_1$
   (Equation~\ref{eq-chione}) values for 1,000 MC data sets
   in the bin located at 5 hours RA, 40$^\circ$ DEC.}
\end{figure}

The bin-by-bin distributions of $\chi_1$ are non-Gaussian
(Figure~\ref{distrib}) and vary as a function of position in the
sky. Thus it is necessary to develop a technique to evaluate the
significance of possible sources. Our technique uses the $\chi_1$
information in neighboring bins to pick out significant
fluctuations above background from the skymap. The parameters in
the technique are tuned on simulated point-like sources.

\clearpage

\section{The Monte Carlo; Simulation of Point-Like Sources}
\label{sec-simsrc}

We have two objectives in simulating point-like sources: The first
is using these simulated sources to tune point source selection
criteria. Secondly, simulated sources provide a straighforward
method by which to quantify our sensitivity to point-sources and
derive flux upper limits.

Simulated source skymaps are created by randomly replacing events
in a simulated isotropic data set with $N_S$ events at the chosen
position for the source. The central-value coordinates of the
simulated source event are randomly shifted according to the error
ellipse, which is taken from the replaced event. The shift
simulates the effect of detector resolution only.  Finally, the
orientations of error ellipses are randomized.

\begin{figure}[h]
   \epsfxsize=18.0cm
   \centerline{\epsffile{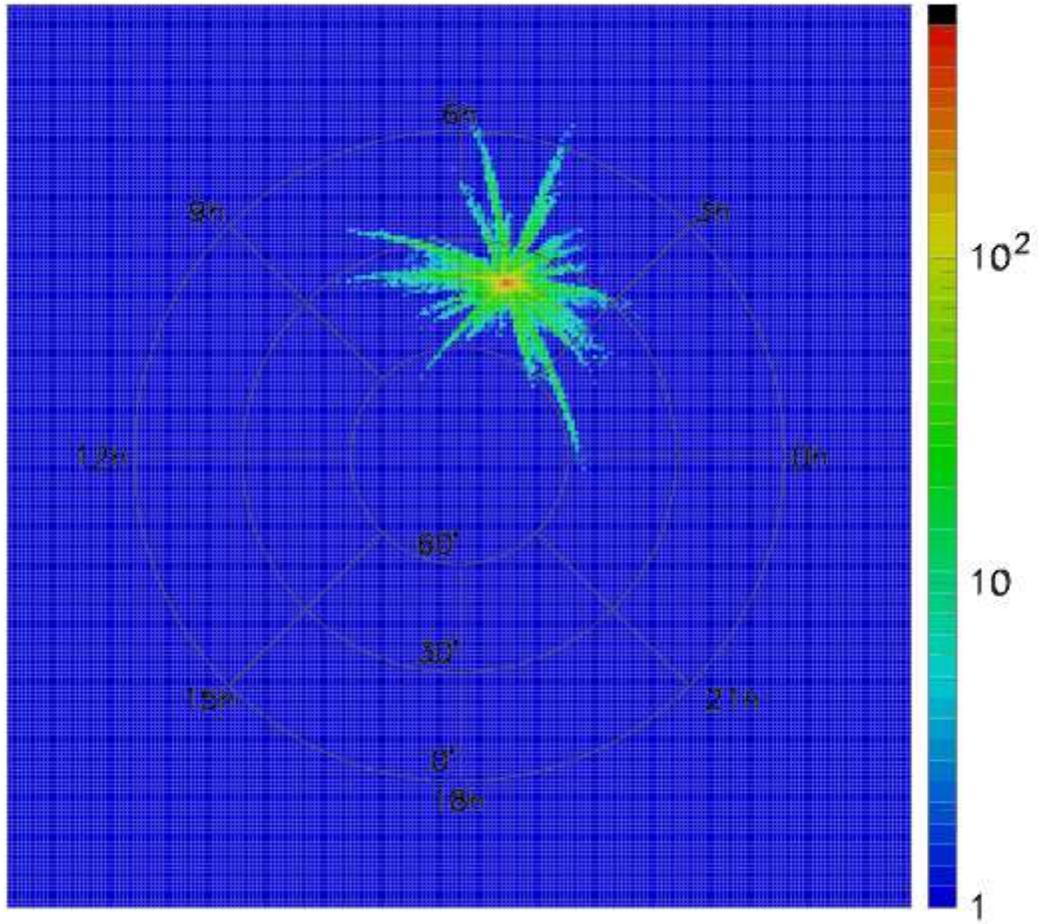}}
   \caption{\label{srconly} An $N_S = 25$ event source shown
   without the isotropic background. The source has been inserted
   at 5 hours RA, 40$^\circ$ DEC.  Each source event is represented by 1,000 points randomly
   thrown according to the elliptical Gaussian error model of
   Equations~\ref{eq-sdp} and \ref{eq-psi}, where the error ellipse is taken from the replaced isotropic event.}
\end{figure}

\begin{figure}[h]
   \epsfxsize=18.0cm
   \centerline{\epsffile{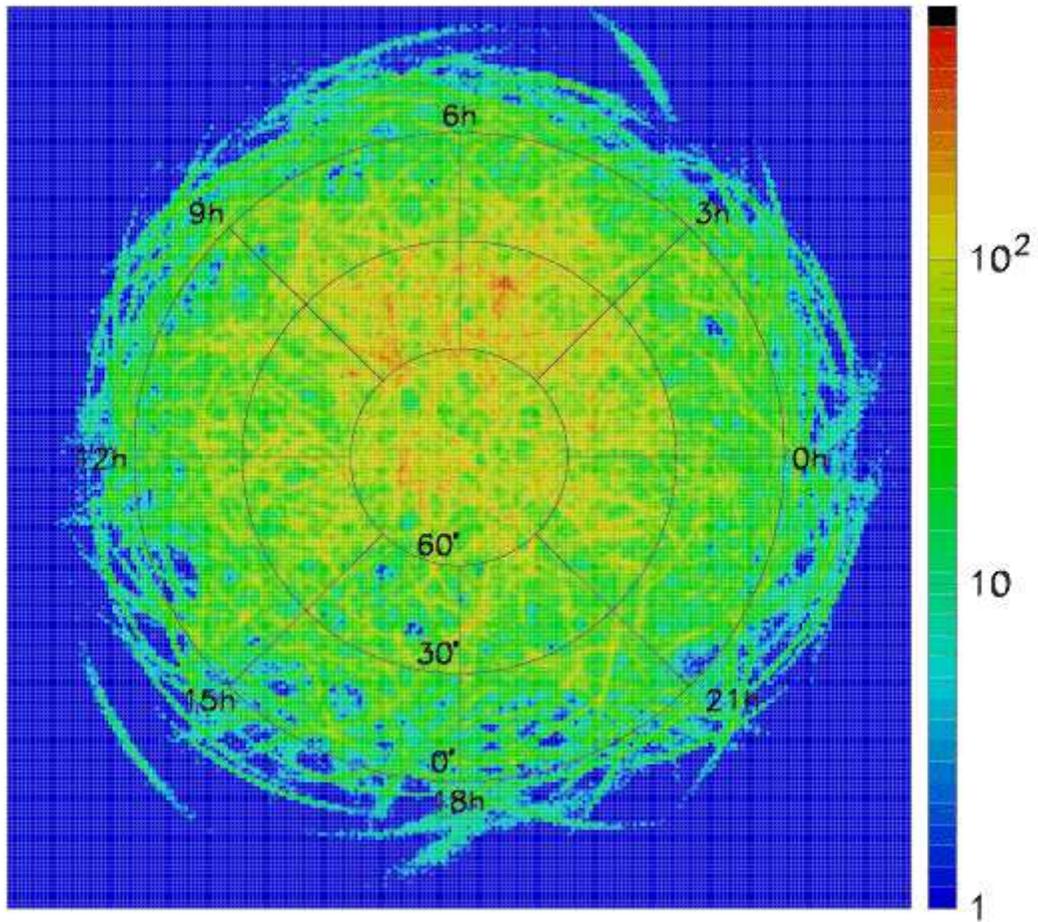}}
   \caption{\label{srconback} Skymap of arrival directions of events
    for a simulated data set, having the same overall exposure as the
    HiRes--I monocular data set, with a 25 event source superimposed at
    5 hours RA, 40$^\circ$ DEC (compare to Figure~\ref{skymap}).}
\end{figure}

\begin{figure}[h]
   \epsfxsize=18.0cm
   \centerline{\epsffile{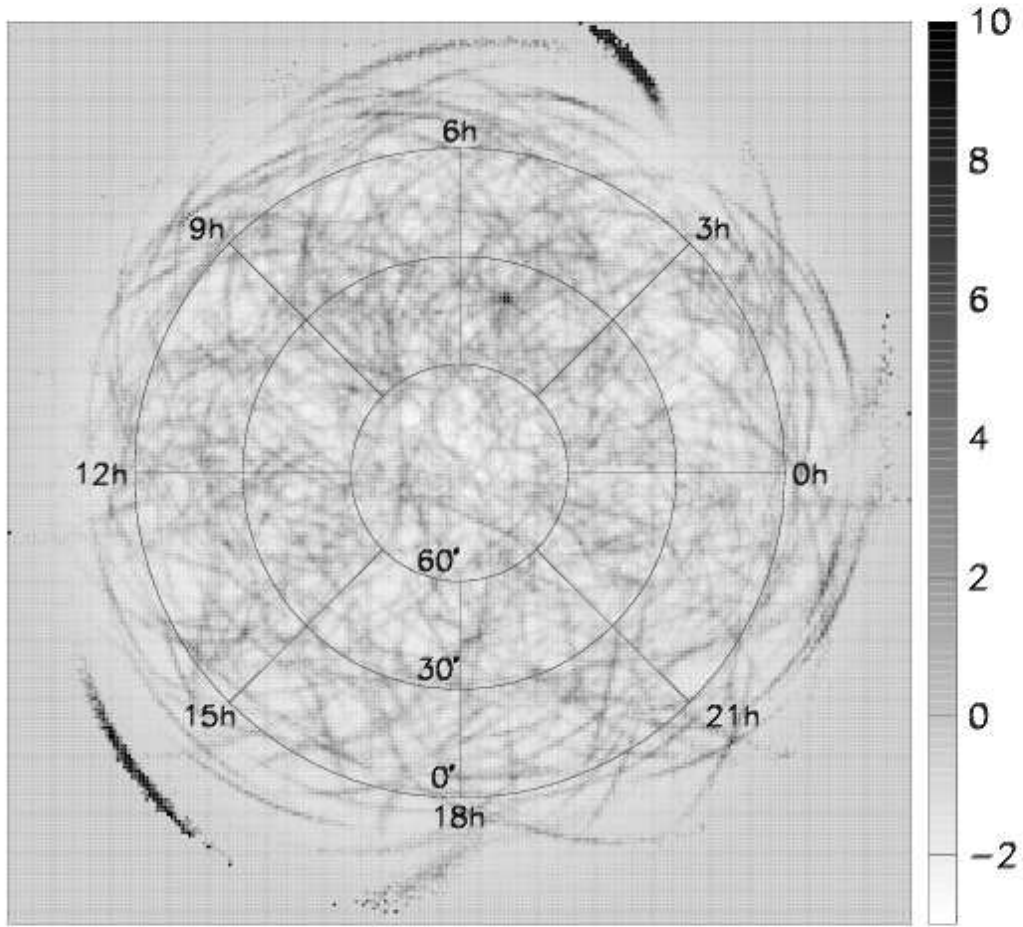}}
   \caption{\label{chionesrc} $\chi_1$ (Equation~\ref{eq-chione}) for
   $N_S=25$ event source inserted in an simulated isotropic data set. The
   source has been inserted at 5 hours RA,
   40$^\circ$ DEC.}
\end{figure}

An example of a simulated source is shown in Figure~\ref{srconly}.
This source is superimposed on a Monte Carlo data set in
Figure~\ref{srconback}, and the quantity $\chi_1$
(Equation~\ref{eq-chione}) is evaluated for each bin in
Figure~\ref{chionesrc}. We note that source events overlap in a
fairly small angular region, as seen in Figure~\ref{srconly}. Thus
we have sensitivity to fairly compact deviations from isotropy, in
spite of our elongated error ellipses.


\clearpage

\section{Calculation of Significances}
\label{sec-significance}

We now describe a procedure by which we can identify point-like
behavior in arrival direction (for example, the simulated source
of Figures \ref{srconly}, \ref{srconback}, and \ref{chionesrc})
while simultaneously rejecting false positives arising from
fluctuations of the background.

Due to detector resolution, it is desirable that we search for
sources by considering points over an extended angular region.
We consider a ``search circle'' of radius $R$, where $R$
is expressed as an angle in degrees. Within the search circle,
we count the fraction of bins $F$ having a $\chi_1$ value
greater than some threshold $\chi_{THR}$. The parameters $R$
and $\chi_{THR}$ are chosen to optimize the signal size, and a cut
is chosen on the fraction $F$ which reduces the false positive
probability to an acceptable level.

Our maximum sensitivity to point-like behavior in arrival
direction, given the HiRes--I pointing uncertainty, was determined
to require a search circle of $R = 2.5^{\circ}$, and a value
$\chi_{THR} = 4$. (In the case in which the bin densities are
normally distributed, this corresponds to 4$\sigma$.)
 The optimum values for these parameters were determined by
simulating sources at various locations in the sky and maximizing
our sensitivity to these sources.  The values for these parameters
are found to be largely insensitive to the position in the sky and
the number of events in the source. Additionally, small variations
in either of these parameters do not have a significant impact on
our results.

Due to low statistics at the edge of HiRes' acceptance, we
consider only search circles with centers whose declinations are
greater than $0^{\circ}$. That is, we only search for sources
north of the celestial equator. Approximately 10\% of HiRes events
have central-value coordinates south of the equator.  These events
can contribute to the search if their error ellipses extend north
of DEC $= -2.5^{\circ}$.

\begin{figure}[h]
   \epsfxsize=18.0cm
   \centerline{\epsffile{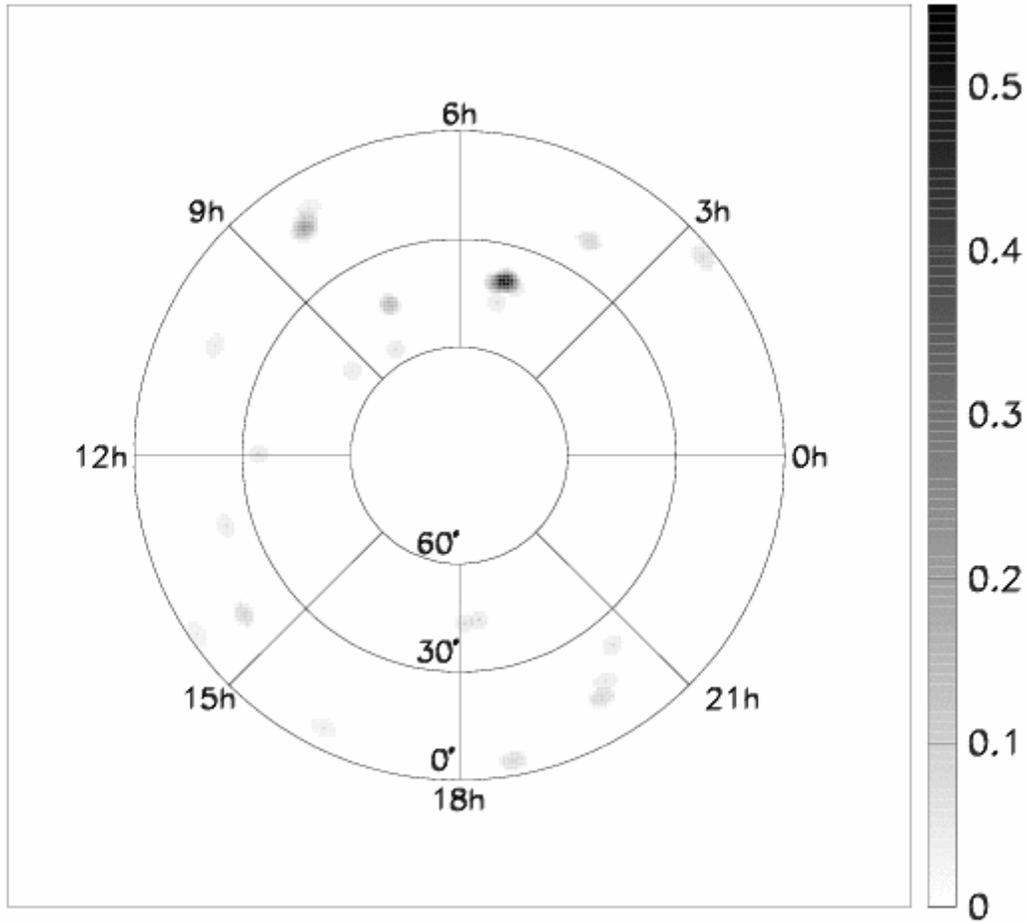}}
   \caption{\label{srcratio} $F$ distribution derived from the
   $\chi_1$ map of Figure~\ref{chionesrc}. $F$ is the fraction
   of bins within radius $R = 2.5^{\circ}$ having a $\chi_1$ value
   of 4 or greater. The simulated source at 5 hours RA, 40$^\circ$ DEC
   clearly stands out as having an exceptionally large excess fraction
   $F$.}
\end{figure}

In Figure~\ref{srcratio}, we have plotted for each bin the fraction
F, for $R = 2.5^{\circ}$ and $\chi_{THR} = 4$, of the
simulated point source of Figure~\ref{srconback}.
The simulated source stands out clearly in this figure.

\begin{figure}[h]
   \epsfxsize=18.0cm
    \centerline{\epsffile{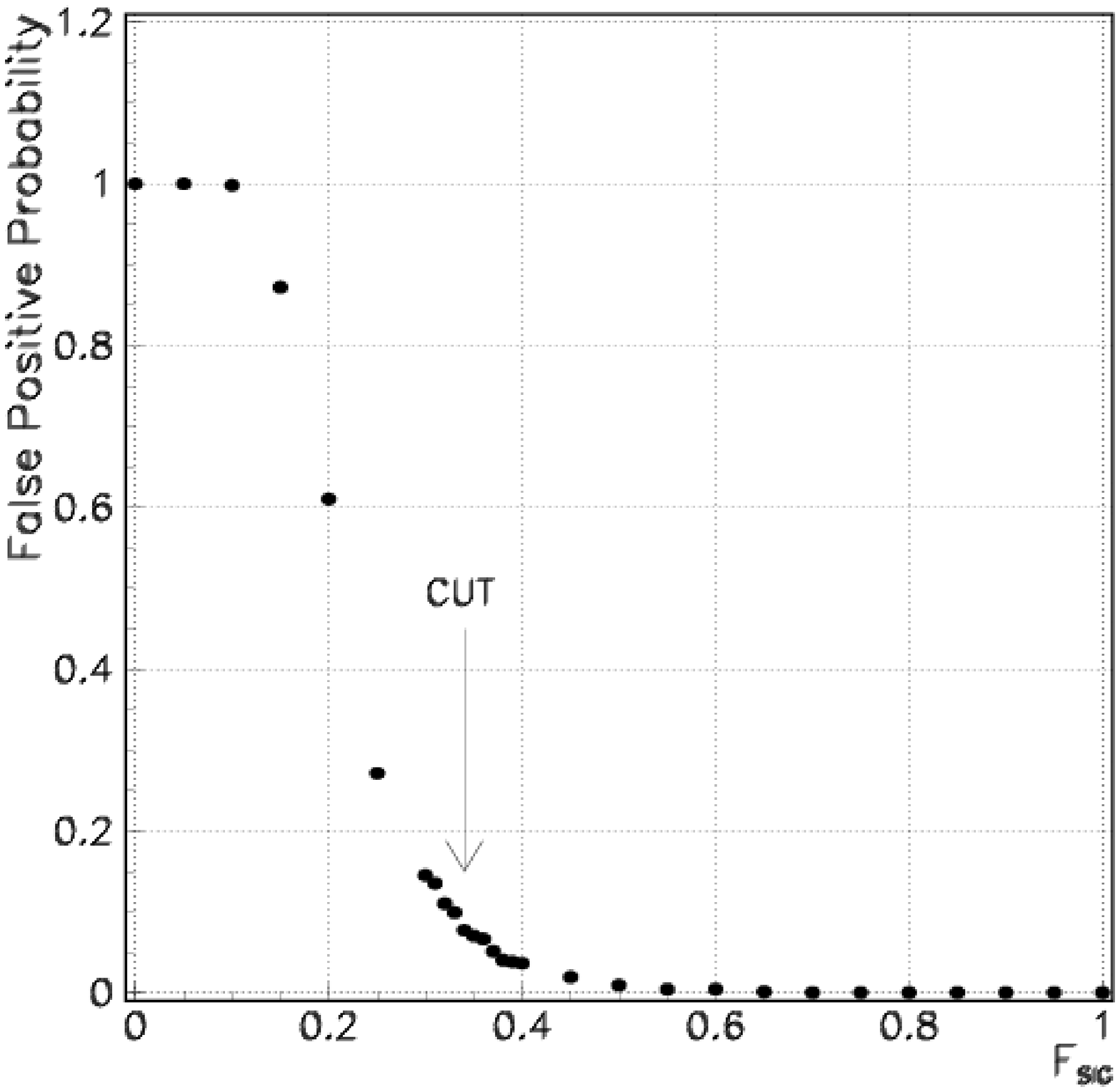}}
   \caption{\label{ratio} Occurrence rate of false positives versus
   $F$, for a $2.5^{\circ}$ search circle and $\chi_1$ threshold
   of 4. A cut at $F = 0.33$ corresponds to a false-positive
   probability of 10\%.}
\end{figure}

The final parameter in this search algorithm is the cut placed on
the quantity $F$.  We evaluate this cut by requiring that the
probability of a simulated isotropic data set -- {\em without} a
superimposed simulated source --- exceeding the cut be no more
than 10\% over the entire sky (Figure~\ref{ratio}). We choose a
cut value of $F = 0.33$, corresponding to a false--positive
probability of 10\%.

\begin{figure}[h]
   \epsfxsize=18.0cm
   \centerline{\epsffile{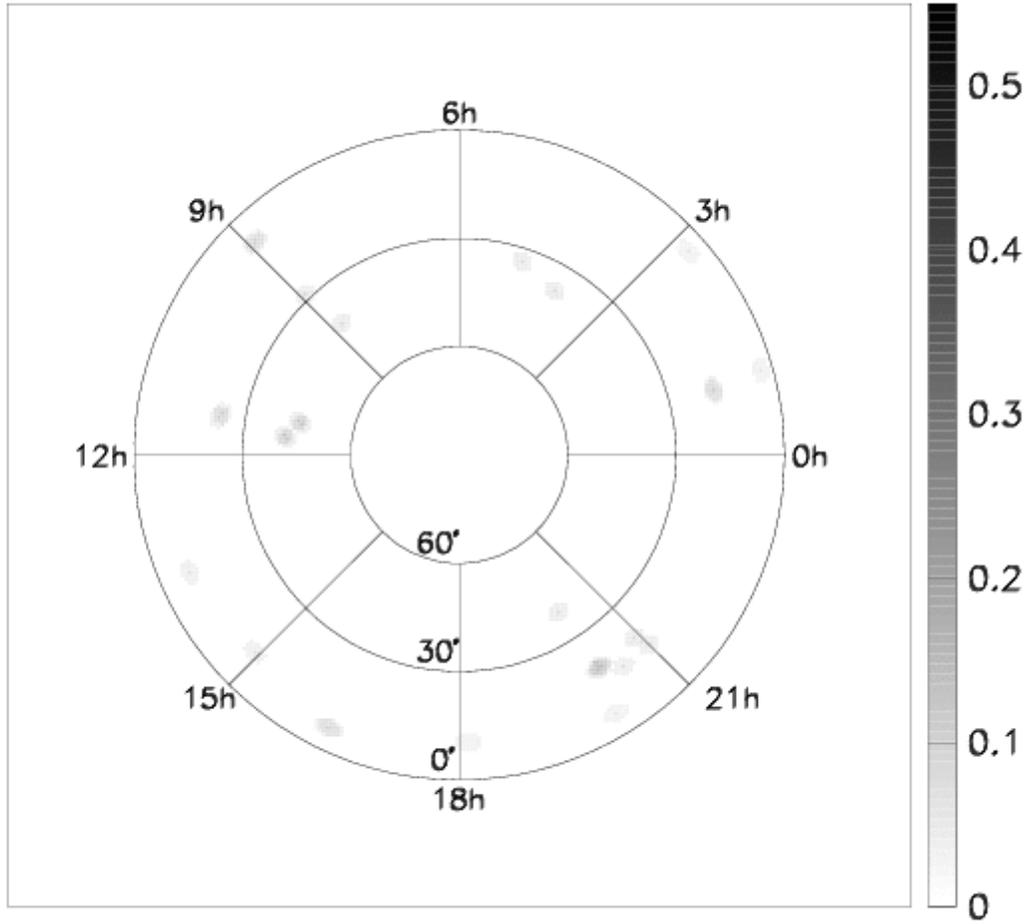}}
   \caption{\label{realratio} $F$ distribution derived from the
   HiRes--I monocular data (Figures~\ref{skymap} and \ref{realchi}).
   $F$ is the fraction of bins within radius $R = 2.5^{\circ}$
   having a $\chi_1$ value of 4 or greater. No points satisfy our search
   criteria. The largest $F$ occurs at DEC $= 20^{\circ}$ and
   RA = 20 hours and has a value $F = 0.15$, corresponding to a
   false-positive probability of 87\%.}
\end{figure}

Figure~\ref{realratio} shows the $F$ distribution for the
monocular data set.  The ``hottest'' spot on this graph, near DEC
$= 20^{\circ}$ and RA = 20 hours, has a value $F = 0.15$.  87\% of
simulated isotropic data sets have $F \geq 0.15$ (see
Figure~\ref{ratio}). We conclude that our observation is
consistent with a fluctuation from an isotropic background.

Next, we evaluate the corresponding sensitivity and flux upper
limits as a function of position in the sky.

\clearpage

\section{Sensitivity and Upper Limits}
\label{sec-limits}

In the preceding section, we found no evidence for the presence of
point-like excesses in the HiRes--I monocular dataset above
$10^{18.5}$ eV. Further, the significance of the largest
point-like fluctuation in the data is well below the threshold
established to minimize the likelihood of false positives in an
isotropic distribution. To quantify our null result, we follow the
suggestion of Feldman and Cousins~\cite{feldman} and calculate
both a set of flux upper limits and the ``sensitivity'' of the
experiment to such point-like excesses. The results of these
calculations are reported below.

\subsection{Sensitivity}

The sensitivity of the experiment is defined as the average 90\%
confidence level flux upper limit that would be reported by an
ensemble of like experiments with no true signal. Since this
average upper limit will vary as a function of position on the
skymap due to different background expectations, we calculate our
sensitivity at set of gridpoints (Table~\ref{gridpts}) distributed
evenly across the Northern Hemisphere. We choose the right
ascension values of our gridpoints to correspond approximately to
the HiRes ``solstices'' and ``equinoxes'', {\em i.e.} to the RA
lines of high/low and midrange event statistics.

To determine our sensitivity to a number of ``source'' events
$\left<N_S\right>$ at a given gridpoint, we generate 400 simulated
isotropic datasets with point-like sources superimposed. The
number of events in each source is Poisson distributed with mean
value $\left<N_S\right>$. These datasets contain the same total
number of events as the HiRes-I data.

We then determine the percentage of trials at each location for
which our reconstruction algorithm ``finds'' a source of size
$\left<N_S\right>$. In the case of the sensitivity calculation, we
say the algorithm ``finds'' a source if at one point on the skymap
$F$ fluctuates above our preselected threshold value of $F=0.33$.
The value of $\left<N_S\right>$ for which signal was declared for
90\% or better of the trials was termed $N_{.33}$.

\begin{figure}[h]
    \epsfxsize=12.0cm
    \centerline{\epsffile{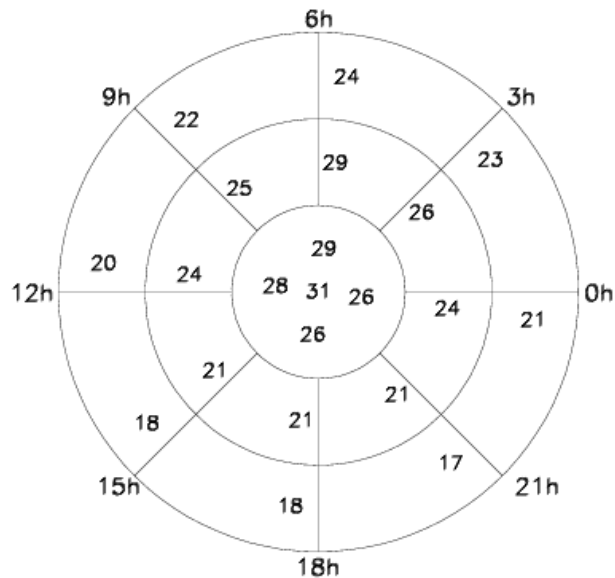}}
    \caption{\label{n33} Numerical values of $N_{.33}$ ---
    the mean number of source events for which signal was declared in
    90\% or better of 400 trials with a cut on F of .33 --- at 21
    grid points in the Northern Hemisphere. Exact numbers and locations
    of gridpoints are given in Table~\ref{gridpts}. The systematic
    uncertainty in the calculation of $N_{.33}$, due to uncertainties
    in the size of the error ellipses, is $\leq$ 1 event.}
\end{figure}

The distribution of $N_{.33}$ at our grid points is illustrated in
Figure~\ref{n33}. We estimate that the systematic uncertainty in
the calculation of $N_{.33}$, due to uncertainties in the size of
the error ellipses, is $\leq$ 1 event.

\begin{figure}[h]
    \epsfxsize=12.0cm
    \centerline{\epsffile{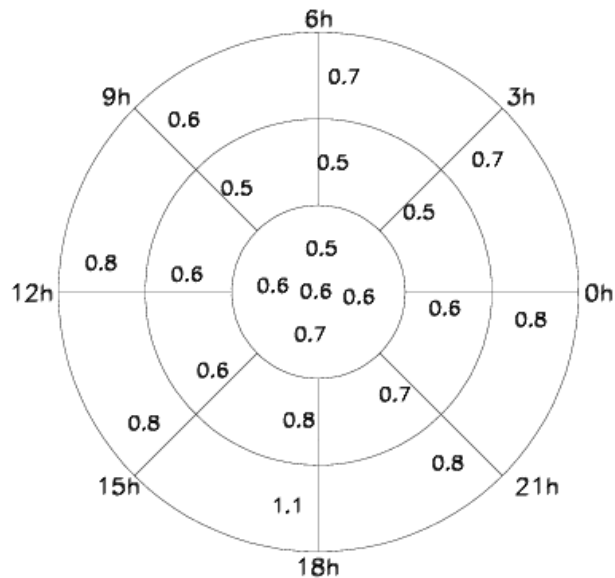}}
   \caption{\label{sensitivity} Flux sensitivity (cosmic rays/km$^2$yr) at
   21 grid points in the Northern Hemisphere. Sensitivity is calculated
   by dividing $N_{.33}$ (Figure~\ref{n33}) by the exposure at each grid
   point. The numbers along with the exact locations of grid points are
   given in Table~\ref{gridpts}.}
\end{figure}

The HiRes-I detector flux sensitivity at each grid point is
$N_{.33}$ at that point divided by the local exposure. We
calculate the detector exposure~\cite{ben-thesis} for point
sources at the grid points by the following procedure: Monte Carlo
events are generated at the grid points, assigned a time from the
distribution of HiRes detector ontimes, and projected towards the
detector aperture. Local coordinates and times are determined,
then the event is paired with a shower from the Monte Carlo event
library having similar local coordinates. An attempt is then made
to reconstruct the Monte Carlo event with the profile--constrained
fitting technique. The exposure, defined as the fraction of events
reconstructed multiplied by the detector aperture (area) and time,
can then be used to determine flux sensitivity (as well as flux
upper limits) for each of the grid locations. These exposures are
listed in Table~\ref{gridpts}. The final flux sensitivities are
shown in Table~\ref{gridpts} and Figure~\ref{sensitivity}.

\subsection{Upper Limits}

\begin{figure}[h]
    \epsfxsize=12.0cm
    \centerline{\epsffile{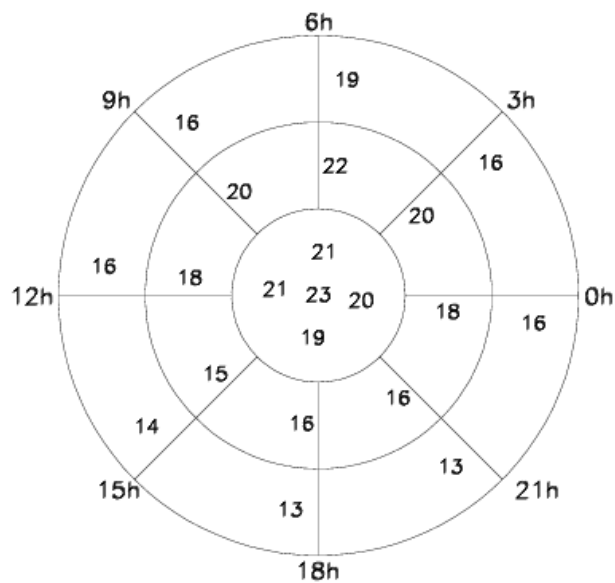}}
   \caption{\label{n15} Numerical values of $N_{.15}$ --- the mean number
   of source events for which signal was declared in 90\% or better of
   400 trials with a cut on $F = .15$ --- at 21 grid points in the Northern
   Hemisphere. Exact numbers and locations of gridpoints are given in
   Table~\ref{gridpts}.The systematic uncertainty in the calculation of
   $N_{.15}$, due to uncertainties in the size of the error ellipses,
   is $\leq$ 1 event.}
\end{figure}

\begin{figure}[h]
    \epsfxsize=12.0cm
    \centerline{\epsffile{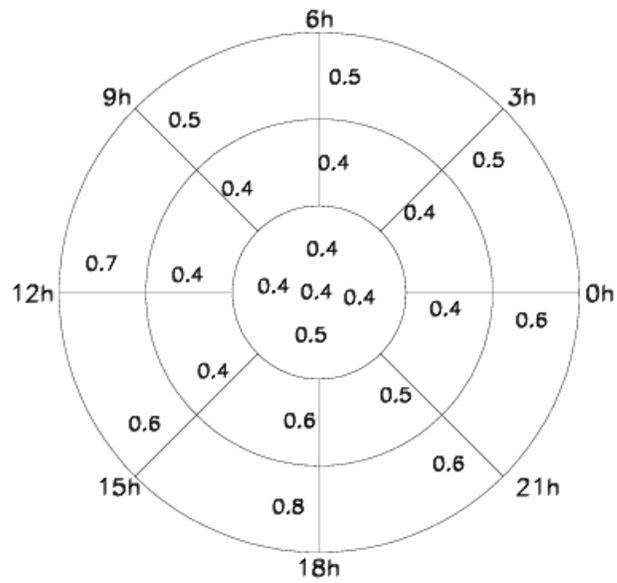}}
   \caption{\label{fluxlimit} 90\% c.l. Flux Upper Limit (cosmic rays/km$^2$yr)
   at 21 grid points in the Northern Hemisphere. Flux upper limit is calculated
   by dividing $N_{.15}$ by the exposure at each grid point.  Exact numbers
   and locations of gridpoints are given in Table~\ref{gridpts}.}
\end{figure}

We place our 90\% confidence level flux upper limits by making use
of the fact that $F$ never fluctuated above 0.15 in the HiRes-I
data. (Figure~\ref{srcratio}. We determine the value of $N_{.15}$
--- the mean number of source events for which signal was declared
in 90\% or better of trials with a threshold of  $F = .15$ --- in
the same manner in which we determined $N_{.33}$ in the previous
section. he results are summarized in Figure~\ref{n15}. Our 90\%
c.l. flux upper limit at each grid point (Figure~\ref{fluxlimit})
is $N_{.15}$ at that point divided by the local exposure. The
largest flux upper limit across the entire sky is 0.8 cosmic rays
above $10^{18.5}$eV per km$^2$yr.

Finally, we note that the {\em a priori} source candidate in the
direction of Cygnus X-3 (RA 20.5 hours, DEC 40.7$^{\circ}$) is
very near the grid point located at RA 20.5 hours, DEC
45$^{\circ}$. We place a 90\% c.l. flux upper limit from Cygnus
X-3 at 0.5 cosmic rays above $10^{18.5}$eV  per km$^2$yr. Previous
Cygnus X--3 flux results were drawn from events samples with energies
above $5 \times 10^{17}$ eV, so a direct comparison is impossible.
We can infer --- assuming that any cosmic rays from Cygnus X-3 have
an energy spectrum similar to that of the full sky --- that an
extrapolation of our result is not competitive with prior upper
limits for events above $5 \times 10^{17}$ eV. However, this is
the first reported measurement for a high-statistics sample above
$10^{18.5}$eV.

\begin{table}[h]
\begin{center}
\begin{tabular}{|c|c|c|c|c|c|c|}
\hline
DEC & RA & & & Exposure & Sensitivity & Upper Limit\\
($deg$) & ($hours$) & \raisebox{1.5ex}{$N_{.33}$} & \raisebox{1.5ex}{$N_{.15}$} & ($km^2yr$) & (cosmic rays/km$^2$yr) & (cosmic rays/km$^2$yr)\\[.5ex]
\hline\hline
&  2.5 hrs & 23 & 16 & 34.2 & .7 & .5\\
\cline{2-7}
&  5.5 hrs & 24 & 19 & 36.6 & .7 & .5\\
\cline{2-7}
&  8.5 hrs & 22 & 16 & 34.3 & .6 & .5\\
\cline{2-7}
& 11.5 hrs & 20 & 16 & 24.5 & .8 & .7\\
\cline{2-7}
\raisebox{1.5ex}{15$^\circ$} & 14.5 hrs & 18 & 14 & 21.9 & .8 & .6\\
\cline{2-7}
& 17.5 hrs & 18 & 13 & 16.7 & 1.1 & .8\\
\cline{2-7}
& 20.5 hrs & 17 & 13 & 21.1 & .8 & .6\\
\cline{2-7}
& 23.5 hrs & 21 & 16 & 26.7 & .8 & .6\\
\hline
&  2.5 hrs & 26 & 20 & 49.7 & .5 & .4\\
\cline{2-7}
&  5.5 hrs & 29 & 22 & 56.6 & .5 & .4\\
\cline{2-7}
&  8.5 hrs & 25 & 20 & 48.5 & .5 & .4\\
\cline{2-7}
& 11.5 hrs & 24 & 18 & 41.2 & .6 & .4\\
\cline{2-7}
\raisebox{1.5ex}{45$^\circ$} & 14.5 hrs & 21 & 15 & 33.5 & .6 & .4\\
\cline{2-7}
& 17.5 hrs & 21 & 16 & 24.9 & .8 & .6\\
\cline{2-7}
& 20.5 hrs & 21 & 16 & 30.3 & .7 & .5\\
\cline{2-7}
& 23.5 hrs & 24 & 18 & 41.5 & .6 & .4\\
\hline
&  5.5 hrs & 29 & 21 & 59.8 & .5 & .4\\
\cline{2-7}
& 11.5 hrs & 28 & 21 & 50.5 & .6 & .4\\
\cline{2-7}
\raisebox{1.5ex}{75$^\circ$} & 17.5 hrs & 26 & 19 & 38.6 & .7 & .5\\
\cline{2-7}
& 23.5 hrs & 26 & 20 & 47.2 & .6 & .4\\
\hline
90$^\circ$ & N/A & 31 & 23 & 53.8 & .6 & .4\\
\hline
\end{tabular}
\vspace{0.5cm} \caption{\label{gridpts} Locations of gridpoints,
threshold signal strengths
         $N_{.33}$ and $N_{.15}$, exposures (with uncertainty 5\%, primarily
         from Monte Carlo statistics), detector flux sensitivity, and 90\%
         confidence level flux upper limits, for cosmic rays with energy
         exceeding $10^{18.5}$~eV.}
\end{center}
\end{table}

\clearpage

\section{Conclusions}
\label{sec-conclusion}

We have conducted a search for point-like excesses in the arrival
direction of ultra--high energy cosmic rays with energy exceeding
$10^{18.5}$ eV in the northern hemisphere. We place an upper limit
of 0.8 cosmic rays/(km$^2$ yr) (90\% c.l.) on the flux from such
sources across the entire sky and place more stringent limits as a
function of position. We also determine sensitivity as a function
of position in the sky.  The HiRes--I monocular data is thus
consistent with the null hypothesis for point-like excesses of
neutral primary cosmic rays in this energy range.

This work is supported by US NSF grants PHY-9321949 PHY-9974537,
PHY-9904048, PHYS-0245428, PHY-0140688, by the DOE grant
FG03-92ER40732, and by the Australian Research Council.
M.A.K. acknowledges the support of a Montana Space Grant Consortium
Fellowship. We gratefully acknowledge the contributions from the
technical staffs of our home institutions and the Utah Center for
High Performance Computing. The cooperation of Colonels E.~Fischer
and G.~Harter, the US Army, and the Dugway Proving Ground staff is
greatly appreciated.

\end{document}